\documentclass[aps,prl,reprint,superscriptaddress]{revtex4-1}

\usepackage{graphicx}
\usepackage{color}

\begin{document}


\title{Complete event-by-event  $\alpha$/$\gamma(\beta)$ separation in a full-size TeO$_2$ CUORE bolometer by Neganov-Luke-magnified light detection}

\author{L.~Berg\'e}
\affiliation{CSNSM, Univ. Paris-Sud, CNRS/IN2P3, Universit\'e Paris-Saclay, 91405 Orsay, France}

\author{M.~Chapellier}
\affiliation{CSNSM, Univ. Paris-Sud, CNRS/IN2P3, Universit\'e Paris-Saclay, 91405 Orsay, France}

\author{M.~de~Combarieu}
\affiliation{IRAMIS, CEA, Universit\'{e} Paris-Saclay, F-91191 Gif-sur-Yvette, France}

\author{L.~Dumoulin}
\affiliation{CSNSM, Univ. Paris-Sud, CNRS/IN2P3, Universit\'e Paris-Saclay, 91405 Orsay, France}

\author{A.~Giuliani}
\affiliation{CSNSM, Univ. Paris-Sud, CNRS/IN2P3, Universit\'e Paris-Saclay, 91405 Orsay, France}
\affiliation{DISAT, Universit\`a dell'Insubria, 22100 Como, Italy}

\author{M.~Gros}
\affiliation{IRFU, CEA, Universit\'{e} Paris-Saclay, F-91191 Gif-sur-Yvette, France}

\author{P.~de~Marcillac}
\affiliation{CSNSM, Univ. Paris-Sud, CNRS/IN2P3, Universit\'e Paris-Saclay, 91405 Orsay, France}

\author{S.~Marnieros}
\affiliation{CSNSM, Univ. Paris-Sud, CNRS/IN2P3, Universit\'e Paris-Saclay, 91405 Orsay, France}

\author{C.~Nones}
\affiliation{IRFU, CEA, Universit\'{e} Paris-Saclay, F-91191 Gif-sur-Yvette, France}

\author{V.~Novati}
\affiliation{CSNSM, Univ. Paris-Sud, CNRS/IN2P3, Universit\'e Paris-Saclay, 91405 Orsay, France}

\author{E.~Olivieri}
\affiliation{CSNSM, Univ. Paris-Sud, CNRS/IN2P3, Universit\'e Paris-Saclay, 91405 Orsay, France}

\author{B.~Paul}
\affiliation{IRFU, CEA, Universit\'{e} Paris-Saclay, F-91191 Gif-sur-Yvette, France}

\author{D.V.~Poda}
\affiliation{CSNSM, Univ. Paris-Sud, CNRS/IN2P3, Universit\'e Paris-Saclay, 91405 Orsay, France}
\affiliation{Institute for Nuclear Research, 03028 Kyiv, Ukraine}

\author{T.~Redon}
\affiliation{CSNSM, Univ. Paris-Sud, CNRS/IN2P3, Universit\'e Paris-Saclay, 91405 Orsay, France}

\author{B.~Siebenborn}
\affiliation{Karlsruhe Institute of Technology, Institut f\"{u}r Kernphysik, 76021 Karlsruhe, Germany}

\author{A.S.~Zolotarova}
\affiliation{IRFU, CEA, Universit\'{e} Paris-Saclay, F-91191 Gif-sur-Yvette, France}


\author{E.~Armengaud}
\affiliation{IRFU, CEA, Universit\'{e} Paris-Saclay, F-91191 Gif-sur-Yvette, France}

\author{C.~Augier}
\affiliation{Univ Lyon, Universit\'{e} Lyon 1, CNRS/IN2P3, IPN-Lyon, F-69622, Villeurbanne, France}

\author{A.~Beno\^{\i}t}
\affiliation{CNRS-N\'eel, 38042 Grenoble Cedex 9, France}

\author{J.~Billard}
\affiliation{Univ Lyon, Universit\'{e} Lyon 1, CNRS/IN2P3, IPN-Lyon, F-69622, Villeurbanne, France}

\author{A.~Broniatowski}
\affiliation{CSNSM, Univ. Paris-Sud, CNRS/IN2P3, Universit\'e Paris-Saclay, 91405 Orsay, France}
\affiliation{Karlsruhe Institute of Technology, Institut f\"{u}r Experimentelle Teilchenphysik, 76128 Karlsruhe, Germany}

\author{P.~Camus}
\affiliation{CNRS-N\'eel, 38042 Grenoble Cedex 9, France}

\author{A.~Cazes}
\affiliation{Univ Lyon, Universit\'{e} Lyon 1, CNRS/IN2P3, IPN-Lyon, F-69622, Villeurbanne, France}

\author{F.~Charlieux}
\affiliation{Univ Lyon, Universit\'{e} Lyon 1, CNRS/IN2P3, IPN-Lyon, F-69622, Villeurbanne, France}

\author{M.~De~Jesus}
\affiliation{Univ Lyon, Universit\'{e} Lyon 1, CNRS/IN2P3, IPN-Lyon, F-69622, Villeurbanne, France}

\author{K.~Eitel}
\affiliation{Karlsruhe Institute of Technology, Institut f\"{u}r Kernphysik, 76021 Karlsruhe, Germany}

\author{N.~Foerster}
\affiliation{Karlsruhe Institute of Technology, Institut f\"{u}r Experimentelle Teilchenphysik, 76128 Karlsruhe, Germany}

\author{J.~Gascon}
\affiliation{Univ Lyon, Universit\'{e} Lyon 1, CNRS/IN2P3, IPN-Lyon, F-69622, Villeurbanne, France}


\author{Y.~Jin}
\affiliation{Laboratoire de Photonique et de Nanostructures, CNRS, Route de Nozay, 91460 Marcoussis, France} 

\author{A.~Juillard}
\affiliation{Univ Lyon, Universit\'{e} Lyon 1, CNRS/IN2P3, IPN-Lyon, F-69622, Villeurbanne, France}

\author{M.~Kleifges}
\affiliation{Karlsruhe Institute of Technology, Institut f\"{u}r Prozessdatenverarbeitung und Elektronik, 76021 Karlsruhe, Germany}

\author{V.~Kozlov}
\affiliation{Karlsruhe Institute of Technology, Institut f\"{u}r Experimentelle Teilchenphysik, 76128 Karlsruhe, Germany}

\author{H.~Kraus}
\affiliation{Department of Physics, University of Oxford, Oxford OX1 3RH, UK}

\author{V.A.~Kudryavtsev}
\affiliation{Department of Physics and Astronomy, University of Sheffield, Hounsfield Road, Sheffield S3 7RH, UK} 

\author{H.~Le~Sueur}
\affiliation{CSNSM, Univ. Paris-Sud, CNRS/IN2P3, Universit\'e Paris-Saclay, 91405 Orsay, France}

\author{R. Maisonobe}
\affiliation{Univ Lyon, Universit\'{e} Lyon 1, CNRS/IN2P3, IPN-Lyon, F-69622, Villeurbanne, France}

\author{X.-F.~Navick}
\affiliation{IRFU, CEA, Universit\'{e} Paris-Saclay, F-91191 Gif-sur-Yvette, France}

\author{P.~Pari}
\affiliation{IRAMIS, CEA, Universit\'{e} Paris-Saclay, F-91191 Gif-sur-Yvette, France}

\author{E.~Queguiner}
\affiliation{Univ Lyon, Universit\'{e} Lyon 1, CNRS/IN2P3, IPN-Lyon, F-69622, Villeurbanne, France}

\author{S.~Rozov}
\affiliation{Laboratory of Nuclear Problems, JINR, 141980 Dubna, Moscow region, Russia } 

\author{V.~Sanglard}
\affiliation{Univ Lyon, Universit\'{e} Lyon 1, CNRS/IN2P3, IPN-Lyon, F-69622, Villeurbanne, France}


\author{L.~Vagneron}
\affiliation{Univ Lyon, Universit\'{e} Lyon 1, CNRS/IN2P3, IPN-Lyon, F-69622, Villeurbanne, France}

\author{M.~Weber}
\affiliation{Karlsruhe Institute of Technology, Institut f\"{u}r Prozessdatenverarbeitung und Elektronik, 76021 Karlsruhe, Germany}

\author{E.~Yakushev}
\affiliation{Laboratory of Nuclear Problems, JINR, 141980 Dubna, Moscow region, Russia }

\begin{abstract}
In the present work, we describe the results obtained with a large ($\approx 133$~cm$^3$) TeO$_2$ bolometer, with a view to a search for neutrinoless double-beta decay ($0\nu\beta\beta$) of $^{130}$Te. We demonstrate an efficient $\alpha$ particle discrimination (99.9\%) with a high acceptance of the $0\nu\beta\beta$ signal (about 96\%), expected at $\approx 2.5$~MeV. This unprecedented result was possible thanks to the superior performance (10~eV rms baseline noise) of a Neganov-Luke-assisted germanium bolometer used to detect a tiny (70~eV) light signal from the TeO$_2$ detector, dominated by $\gamma$($\beta$)-induced Cherenkov radiation but exhibiting also a clear scintillation component. The obtained results represent a major breakthrough towards the TeO$_2$-based version of CUORE Upgrade with Particle IDentification (CUPID), a ton-scale cryogenic $0\nu\beta\beta$ experiment proposed as a follow-up to the CUORE project with particle identification. The CUORE experiment began recently a search for neutrinoless double-beta decay of $^{130}$Te with an array of 988 125-cm$^3$~TeO$_2$ bolometers. The lack of $\alpha$ discrimination in CUORE makes $\alpha$ decays at the detector surface the dominant background component, at the level of $\approx 0.01$~counts/(keV~kg~y) in the region of interest. We show here, for the first time with a CUORE-size bolometer and using the same technology as CUORE for the readout of both heat and light signals, that surface $\alpha$  background can be fully rejected.

\end{abstract} 
\maketitle

The observation of the neutrinoless double-beta decay ($0\nu2\beta$) would have deep implications for our understanding of nature. It would signal the breaking of the lepton number symmetry, ascertain that neutrinos are Majorana fermions, and provide a neat mechanism to explain the smallness of neutrino masses and the abundance of matter over antimatter in the present Universe \cite{Vergados:2016,Pas:2015}. This rare nuclear process consists of a transformation of an even-even nucleus into a lighter isobar containing two more protons and accompanied by the emission of two electrons and no neutrinos, violating the lepton number conservation. Although $0\nu2\beta$ is crucial to explore the fundamental neutrino nature, searches for this transition are much more than neutrino-physics experiments. In fact, $0\nu2\beta$ is a powerful, inclusive test of lepton number violation (LNV) in the nuclear matter. LNV is as important as baryon number violation in theories beyond the Standard Model \cite{Pas:2015}. No evidence of the $0\nu 2\beta$ existence has been reported over around 70 years of experimental searches, and the most stringent half-life limits are in the range $10^{24}$--$10^{26}$~yr \cite{Vergados:2016,Gando:2017}.

\begin{figure*}[t]
\includegraphics[width=0.50\textwidth]{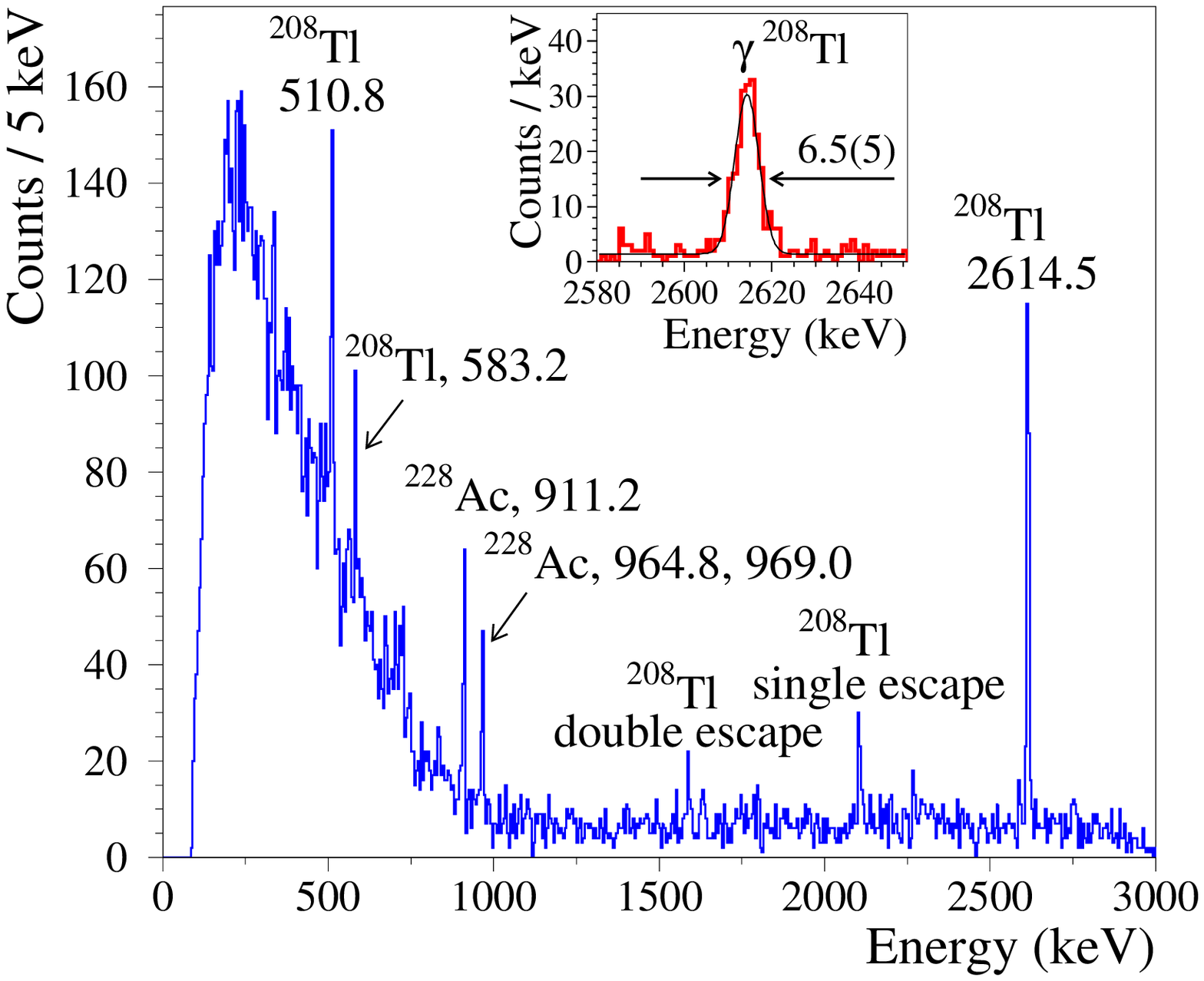}
\includegraphics[width=0.48\textwidth]{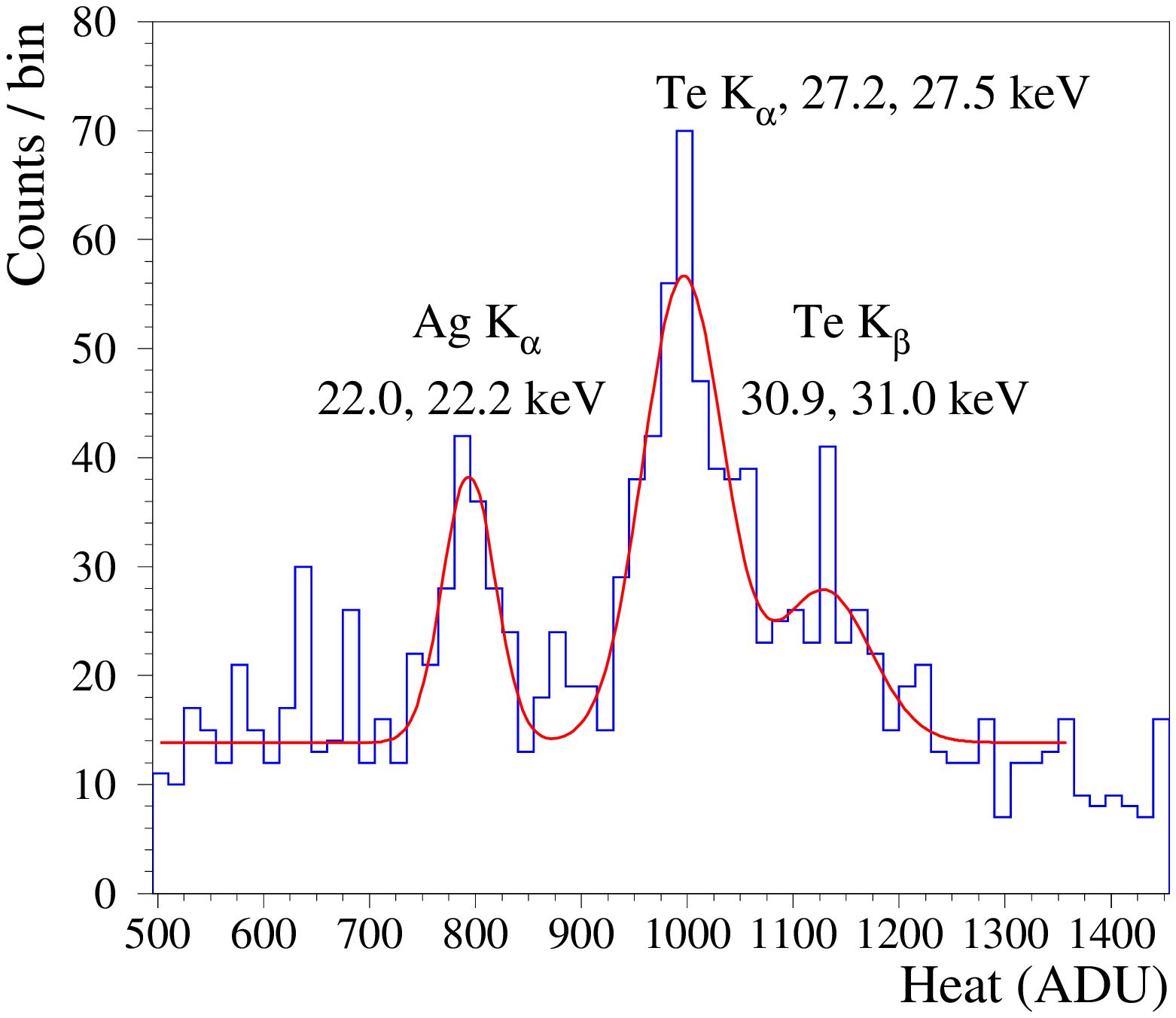}
\caption{Left panel: energy spectrum of a $^{232}$Th $\gamma$ source measured by a 784-g TeO$_2$ bolometer in the EDELWEISS set-up over 148.3~h. The spectrum contains a population of $\alpha$ particles from a smeared $^{210}$Po $\alpha$ source. The energies of the $\gamma$ peaks visible in the spectrum are given in keV. The 2615~keV $\gamma$ peak of $^{208}$Tl, fitted with a Gaussian function and a flat component, is depicted in the inset. The FWHM energy resolution at the peak is 6.5$\pm$0.5~keV.
Right panel: amplitude spectrum of X-rays excited by a $^{60}$Co $\gamma$ source irradiating the holder --- including the TeO$_2$ crystal and the Ag internal layer --- and measured by the Ge light detector operated at 0~V electrode bias (6.4~h of data taking). A fit to the data by a flat distribution and three Gaussian functions is shown by the red curve. The energy scale of the light detector is given in ADU (analog-to-digit units).}
\label{fig:Calibration}
\end{figure*}

The bolometric technique is one of the most powerful methods to investigate $0\nu2\beta$. CUORE (Cryogenic Underground Observatory for Rare Events) studies the decay of the isotope $^{130}$Te via bolometers and is the largest bolometric experiment. It is operated in LNGS (Laboratori Nazionali del Gran Sasso, Italy) and the data taking has just successfully begun \cite{Alduino:2017a}. It will run for several years and will be one of the most sensitive $0\nu2\beta$ experiments of this decade. CUORE is not a background-free experiment: a few tens of background counts per year are expected in the region of interest (ROI) at around 2527 keV, $Q$-value of the $^{130}$Te $0\nu 2\beta$ transition ($Q_{\beta\beta}$). According to the CUORE background model, the dominant component is due to energy-degraded $\alpha$'s, emitted by traces of surface radioactive contamination of the TeO$_2$ crystals and especially of the surrounding copper structures (e.g. see \cite{Artusa:2014,Alduino:2017} and references therein).

The CUPID project (CUORE Upgrade with Particle ID) \cite{CUPID} is the proposed successor of CUORE and aims at improving the sensitivity to the $0\nu2\beta$ half-life by two orders of magnitude, probing the Majorana nature of neutrinos in the so-called inverted hierarchy region of the neutrino mass. To this end, CUPID will rely on bolometers with active background suppression capabilities. One of the most intensive R\&D activities in CUPID \cite{CUPID_RD} is focused on exploiting the Cherenkov light emitted by TeO$_2$. This brilliant method (proposed in Ref.~\cite{Tabarelli:2010}) relies on the fact that $\beta$'s have a Cherenkov light emission threshold of 50 keV, whereas this threshold amounts to 400~MeV for $\alpha$'s, which will deliver no Cherenkov light in the ROI. Up to date, this method has enabled a full $\alpha$/$\gamma$($\beta$) separation only with twice smaller TeO$_2$ bolometers than the CUORE standard element, corresponding approximately to a 5-cm-side cubic crystal. It is to note that the Cherenkov light emitted by the crystal is steeply decreasing with increasing crystal size \cite{Artusa:2016}.\\

In this work we demonstrate, for the first time, a complete $\alpha$/$\gamma$($\beta$) separation with a full-size CUORE TeO$_2$ detector, by coupling the TeO$_2$ crystal to a germanium light detector. The signal of the latter is provided by the same temperature-sensor technology as the TeO$_2$ main bolometer, which coincide with the method adopted in CUORE. Therefore, our technique provides the important advantage of a full compatibility with the current CUORE read-out system. With the active rejection achieved in this work, one can reduce the $\alpha$ background component of CUORE by a factor of $\approx 10^3$ with negligible effects on the $0\nu2\beta$ decay detection efficiency.\\


We used a CUORE-size TeO$_2$ crystal, which belongs to the batch produced for Cuoricino, precursor of the CUORE experiment \cite{Andreotti:2011}. The 51$\times$51$\times$51~mm crystal, with a mass of 784~g, was kept by polytetrafluoroethylene (PTFE) clamps inside a cubic, internally-silver-plated copper housing. A $^{210}$Po $\alpha$ source (made by $^{218}$Po implantation on a copper substrate) was fit on the top cap of the holder. Around 70\% of the $\alpha$ source surface was covered by three stacked 6~$\mu$m-thick Mylar\textsuperscript{\textregistered} foils to degrade the 5.3~MeV $\alpha$ energy and populate the ROI for $^{130}$Te $0\nu2\beta$.

\begin{figure*}[t]
\includegraphics[width=0.49\textwidth]{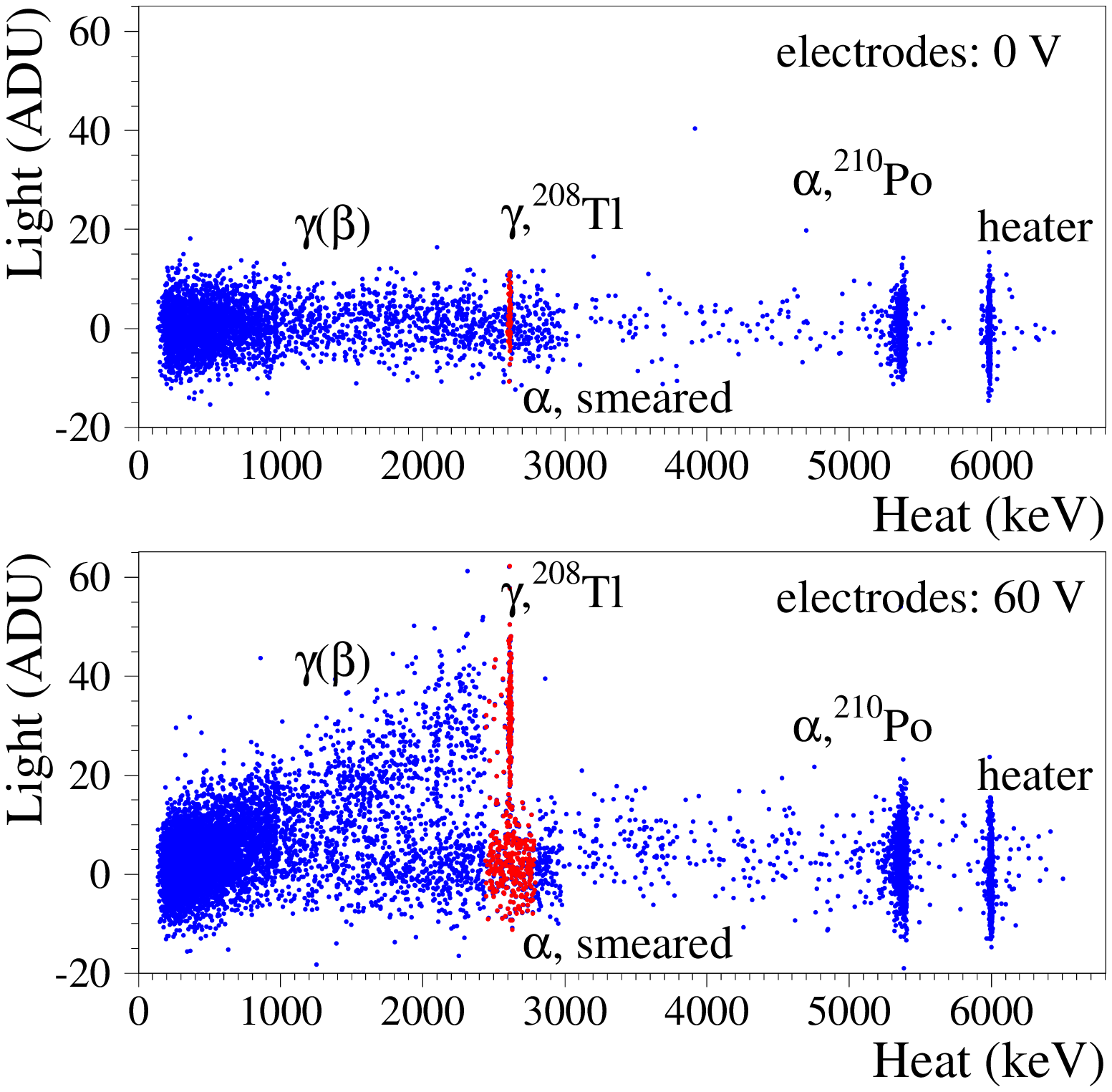}
\includegraphics[width=0.49\textwidth]{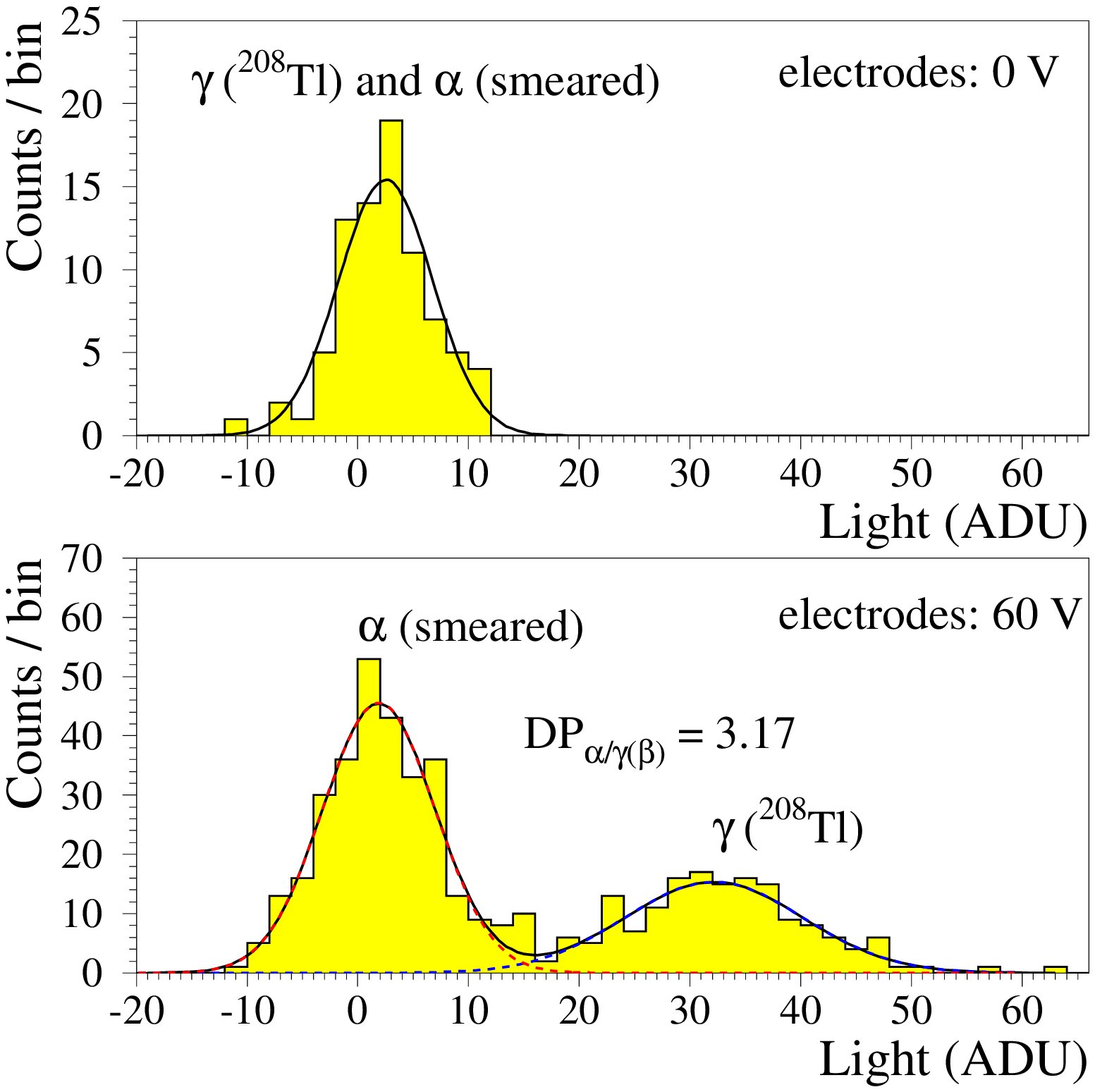}
\caption{Left panel: scatter plots of the heat-vs.-light signals in the measurements with a 784-g TeO$_2$ bolometer in coincidence with a Neganov-Luke Ge light detector operated respectively as a standard light detector (0~V electrode bias -- Top figure) and in a signal-amplification regime (60~V electrode bias -- Bottom figure). The TeO$_2$ detector was irradiated by $\gamma$ quanta from a $^{232}$Th source and $\alpha$ particles from a partially-smeared $^{210}$Po source. Heater events were used for the thermal response stabilization. Right panel: projections onto the y-axis of the events marked in red in the left panel, for 0~V (Top figure) and 60~V (Bottom figure) electrode bias respectively. The light signal distributions are fitted with Gaussian functions. The discrimination power (see text) between $\alpha$ and $\gamma$ events is 3.17. The energy scale of the light detector is given in ADU (analog-to-digit units).}
\label{fig:Separation}
\end{figure*}

The TeO$_2$ crystal side opposite to the $\alpha$ source was exposed to a Neganov-Luke-assisted light detecting bolometer \cite{Neganov:1985,Luke:1988}. The light absorber consists of an electronic-grade Ge wafer ($\oslash$44$\times$0.175~mm) provided with two sets of annular, concentric aluminum electrodes on a side. The electrode geometry allowed us to drift charges parallel to the wafer surface for distances as short as 4~mm, corresponding to the pitch of the annular electrodes. A 70~nm SiO anti-reflecting coating was deposited on the whole surface of the electrode side. Both heat and light detectors were equipped with NTD (Neutron Transmutation Doped, \cite{Haller:1994}) germanium thermistors. These sensors provide, respectively, the heat signals due to particles impinging on the TeO$_2$ bolometer and the light signals associated to the TeO$_2$ out-coming light.
If a voltage potential is applied between two nearby electrodes of the light detector (Neganov-Luke mode), an electric field develops within the germanium wafer. When photons are absorbed by germanium, electrons and holes are created and drifted by the electric field towards the electrodes. Eventually, additional heat is released in the wafer because of the charge motion, magnifying the detector heat signal. A more detailed description of the Neganov-Luke-assisted light detector can be found in Ref.~\cite{Artusa:2017}. A Si:P based heater was glued to the TeO$_2$ crystal with epoxy. It was used to inject periodically a constant energy and thus provide an off-line stabilisation of the detector response \cite{Alessandrello:1998}.\\

The detector ran at 17 mK in the cryostat hosting the dark-matter EDELWEISS experiment in the LSM (Laboratoire Souterrain de Modane, France). The details of the cryogenic facility, the readout electronics and the data acquisition system can be found in Refs. \cite{Armengaud:2017a,Armengaud:2017b}. Data processing was performed with an optimal filter technique \cite{Gatti:1986}. The heat-light coincidences were treated according to Ref. \cite{Piperno:2011}. 

This work reports the results of a week-long calibration measurement with a low activity ($\approx 300$~Bq) $^{232}$Th source placed outside the thermal screens of the cryostat. The energy spectrum of the TeO$_2$ bolometer is shown in Fig.~\ref{fig:Calibration} (Left panel). The detector exhibited excellent performance with heat-energy resolution as good as 6.5(5)~keV FWHM at 2615~keV. The light detector was calibrated by X-ray fluorescence, stimulated in the material surrounding it by the periodical use of an external $\approx 100$~kBq $^{60}$Co source. This calibration was performed for 1~h per day with the electrode voltage bias set to zero. Fig.~\ref{fig:Calibration} (Right panel) shows the light detector calibration spectrum. It returned a baseline noise of 108~eV (rms); this value is not exceptional for an NTD-Ge-based light detectors of this size, for which baseline noises of $\approx 30-50$~eV rms have been already achieved \cite{Artusa:2016}.

A part of the $^{232}$Th measurements (49.6~h) was performed with Neganov-Luke amplification off, $i.e.$ zero electrode bias. A second part of the data (98.7~h) was collected at 60~V electrode bias, at which the detector exhibited the best performance in terms of signal-to-noise ratio. Figure~\ref{fig:Separation} (Left panel) shows the event-by-event heat-light scatter plots for both electrode bias 0~V and 60~V and demonstrates how the Neganov-Luke signal amplification makes the separation between $\gamma$($\beta$) and $\alpha$ possible.

\begin{table*}[t]
\caption{Results of the Gaussian fits (mean value $\mu$ and standard deviation $\sigma$) of the light-signal amplitude distributions corresponding to the baseline, $\gamma$($\beta$) and $\alpha$ particles selected within a given heat energy interval. The data were acquired during a $^{232}$Th calibration of a 784-g TeO$_2$ bolometer coupled to a Neganov-Luke light detector operated at 0~V and 60~V grid bias respectively.} 
\begin{ruledtabular}
\begin{tabular}{c|ccc|ccc}
~ & \multicolumn{3}{c}{0~V grid bias} & \multicolumn{3}{c}{60~V grid bias}\\
\cline{2-7}
~ & Heat (keV) & Light $\mu$ (ADU) & Light $\sigma$ (ADU) & Heat (keV) & Light $\mu$ (ADU) & Light $\sigma$ (ADU) \\
\cline{2-7}
Baseline 					& 0 					& -0.05$\pm$0.05	& 3.92$\pm$0.04	& 0 					& 0.005$\pm$0.06	& 4.54$\pm$0.06 \\
$\gamma$($\beta$) & 2598--2632	& 2.5$\pm$0.5 		& 4.2$\pm$0.3 	& 2440--2790	& 32.1$\pm$0.7		& 8.1$\pm$0.7  \\
$\alpha$ 					& 2640--2790	& 0.1$\pm$0.6			& 4.7$\pm$0.4 	& 2440--2790	& 1.9$\pm$0.3			& 5.1$\pm$0.3  \\
\end{tabular}
\end{ruledtabular}
\label{tab:Fit}
\end{table*} 

The 0~V electrode bias data were used to demonstrate both the lack of $\alpha$/$\gamma$($\beta$) discrimination capability in the case of using a standard light detector and to estimate the light energy output of $\gamma$'s of $^{208}$Tl, selected in a narrow heat energy interval around 2615~keV (see Top plots in Fig.~\ref{fig:Separation}). We will use this information to evaluate the energy sensitivity of the light detector in the Neganov-Luke mode \cite{Artusa:2017}.

Table~\ref{tab:Fit} reports the results of the fits to the light signal distributions corresponding to the baseline noise, the 2615 keV $\gamma$ peak and similar-energy $\alpha$'s (in a 2640--2790~keV interval) detected by the light detector at 0 V electrode bias. The light signal associated to a 2615~keV $\gamma$ is only (70$\pm$13)~eV (corresponding to a light yield of $\approx 27$~eV/MeV), which is around 30\% smaller than what was obtained in previous measurements with TeO$_2$ crystals of similar sizes \cite{Casali:2015,Pattavina:2016}. This is due mainly to the poor geometric coupling between the TeO$_2$ bolometer and the light detector. In particular, the light is transmitted through a hole with around 40\% smaller area than that of the crystal side facing the light detector. This is due to the holder structure imposed by mechanical constraints in the EDELWEISS cryostat. 

We have selected the events in the 2440--2790~keV energy interval (in the vicinity of the $Q_{\beta \beta}$-value of $^{130}$Te) to measure both the light detector performance at 60~V electrode bias and the light-assisted $\alpha$ discrimination capability. We then constructed the corresponding histogram of the light-signal amplitudes (see Bottom plots of Fig.~\ref{fig:Separation}) and fit the distributions of $\alpha$ and $\gamma$($\beta$) events by two Gaussian functions (see Table~\ref{tab:Fit}). 

\begin{figure}
\includegraphics[width=0.45\textwidth]{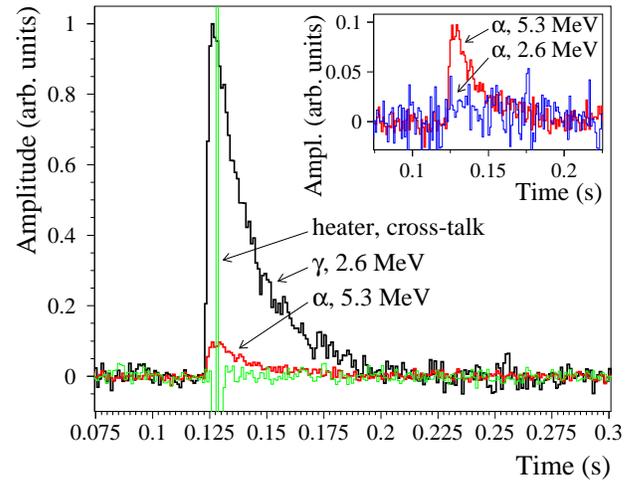}
\caption{Average pulses of a Neganov-Luke light detector acquired in coincidence 
with a 784-g TeO$_2$ bolometer with the following event selection in the main absorber: 2.6 MeV $\gamma$ quanta (333 waveforms), 5.3~MeV $\alpha$ particles (3551 waveforms), and heater-induced cross-talk (1608 waveforms). 
The average light signal induced by 2.6 MeV $\alpha$'s (668 waveforms) is shown in the inset together with the one induced by 5.3~MeV $\alpha$'s. The pulses are normalized at the maximum of the 2.6~MeV $\gamma$ average pulse.}
\label{fig:Meanpulses}
\end{figure}

As one can see in Fig.~\ref{fig:Separation} and Table~\ref{tab:Fit}, the distribution of the light of $\alpha$ events, even below 3 MeV, is not centered around zero,  whereas it is for the heater events, for which neither Cherenkov nor scintillation light is expected. Furthermore, we selected $\approx 2.6$~MeV $\gamma$ and $\alpha$, $\approx 5.3$~MeV $\alpha$, and $\approx 6$~MeV heater events and averaged their individual waveforms. The average pulses, reported in Fig.~\ref{fig:Meanpulses}, unambiguously confirm that TeO$_2$ poorly scintillates at low temperatures, as claimed in \cite{Coron:2004} and hinted by the results of \cite{Beeman:2012,Schaffner:2015,Willers:2015,Pattavina:2016,Gironi:2016,Artusa:2017}. By assuming a quenching factor of $\approx 0.2$ for $\alpha$-induced scintillation, which is a typical value for well-studied crystal scintillators (e.g. see \cite{Armengaud:2017a} and references herein), one could estimate that the TeO$_2$ scintillation contributes by $\approx 20$\% to the 2.6 MeV $\gamma$ light signals. This corresponds to about 5 eV/MeV scintillation light yield, not considering light collection efficiency. This estimation is in a good agreement with room temperature studies of Cherenkov light emission from a TeO$_2$ crystal \cite{Bellini:2012}.  

\begin{figure}
\includegraphics[width=0.45\textwidth]{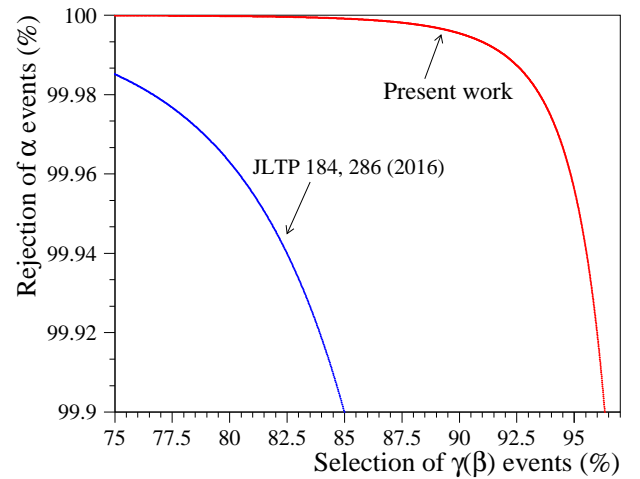}
\caption{The rejection factor of $\alpha$ events versus the acceptance level of $\gamma$($\beta$) 
events based on the results of the $\alpha$/$\gamma$($\beta$) separation in the ROI of $0\nu2\beta$ decay 
of $^{130}$Te achieved with the present work with a 784-g TeO$_2$ bolometer coupled to the Neganov-Luke 
light detector and compared with the only previous test conducted so far with a CUORE-size TeO$_2$ crystal \cite{Pattavina:2016}. The right boundary of the accepted $\gamma$($\beta$) was set at 3$\times \sigma_{Light}$, while the left boundary was varied to get a selection efficiency within 75--97\%. The corresponding $\alpha$ rejection efficiency was computed with an infinite left boundary, while the level of the right boundary was set according to the left boundary of the $\gamma$($\beta$) acceptance band.}
\label{fig:Rejection}
\end{figure} 

By comparing the amplitude of the light of the 2615-keV $\gamma$'s at 0~V and 60~V electrode bias 
(Table~\ref{tab:Fit}), we can estimate a Neganov-Luke amplification factor of 12.7, in agreement with \cite{Artusa:2017}. The light detector baseline noise at 60~V electrode bias is as good as 10$\pm$2~eV (rms), which represents a breakthrough for an optical bolometer of such a large surface, coupled to a CUORE-like TeO$_2$ crystal \cite{Casali:2017,Casali:2015}. 

We define here the $\alpha$/$\gamma$($\beta$) discrimination power, $D_{\alpha/\gamma(\beta)}$, as:

\begin{equation} 
D_{\alpha/\gamma(\beta)} = \frac{\mu_{\gamma(\beta)} - \mu_{\alpha}}{\sqrt{\sigma^{2}_{\gamma(\beta)} + \sigma^{2}_{\alpha}}}.
\label{eq:DP}
\end{equation} 

\noindent According to the fit parameters given in Table \ref{tab:Fit}, the detector showed a $D_{\alpha/\gamma(\beta)}$ of 3.17, value never achieved for a full-size CUORE TeO$_2$ bolometer \cite{Artusa:2017}. In spite of a non-optimal light collection, we report a signal-to-noise ratio of 7.1 for a light signal corresponding to a 2615~keV $\gamma$. This value fulfills the requirement needed to obtain a 3-order-of-magnitude suppression of the $\alpha$ background in CUORE \cite{Casali:2015}. The achieved separation for $\alpha$'s amounts to 5.9~$\sigma$ and complies with the CUPID goal \cite{Beeman:2012, CUPID}. By using the results of the two-Gaussian fits given in Table~\ref{tab:Fit}, we computed the efficiency of $\alpha$-event rejection as a function of the $\gamma$($\beta$) acceptance. Fig.~\ref{fig:Rejection} shows that an $\alpha$ rejection factor of 99.9\% can be achieved with a $\gamma$($\beta$) acceptance of about 96\%, pointing out a dramatic improvement compared to the previous results \cite{Pattavina:2016}.

Summarizing, this work unambiguously demonstrates, for full-size CUORE TeO$_2$ bolometers, an event-by-event active particle identification capability, which complies with the requirements of the CUPID $0\nu 2\beta$ project in terms of $\alpha$ background rejection. We stress that the achieved results were possible only thanks to the superior performance of the Neganov-Luke-assisted light detecting bolometer, whose design and fabrication process have been developed in the CSNSM laboratory (Orsay, France). 
This technology is now mature and provides reproducible results. The separation obtained in this work indicate that it is now ready for deployment in a large-scale bolometric detector such as CUPID.

This work has been partially performed in the framework of the LUMINEU program, a project funded by the Agence Nationale de la Recherche (ANR, France). The help of the technical staff of the Laboratoire Souterrain de Modane and of the other participant laboratories is gratefully acknowledged. We thank the mechanical workshop of CEA/SPEC for its valuable contribution in the conception and fabrication of the detector holders. A.S.Z. is supported by the ``IDI 2015'' project funded by the IDEX Paris-Saclay, ANR-11-IDEX-0003-02.

\bibliography{TeO_GeCo_LSM_PRCr}

\end{document}